\documentclass[12pt]{article}
\usepackage{latexsym}
\usepackage{graphics}
\usepackage{epsfig}
\begin{document}
\pagestyle{plain}
\setcounter{page}{1}
\section*{\noindent The metron model. \\
\noindent Towards a unified deterministic \\
theory of fields
and
particles}
\vspace{5mm}
\subsection*{Klaus Hasselmann\\
\noindent Max Planck Institute for Meteorology, Hamburg}
\vspace{1.2cm}
{\it ABSTRACT}\\

A summary is given of the principal concepts of a unified deterministic
theory of fields and particles that have been developed in more detail in a
previous comprehensive four-part paper (Hasselmann, 1996a,b, 1997a,b). The
model is based on the Einstein vacuum equations, Ricci tensor $R_{LM}=0$,
in a higher-dimensional space. A space of at least eight dimensions is
required to incorporate all other forces as well as  gravity in Einstein's
general relativistic formalism. It is hypothesized that the  equations
support soliton-type solutions ("metrons") that are localized in physical
space and are periodic in extra ("harmonic") space and time.  The solitons
represent waves propagating in harmonic space that are locally trapped in
physical space within a wave guide produced by a distortion of the
background metric. The metric distortion, in turn, is generated by
nonlinear interactions (radiation stresses) of the wave field. (The mutual
interaction mechanism has been demonstrated for a simplified Lagrangian in
Part 1 of the previous paper). In addition to electromagnetic and
gravitational fields, the metron solutions carry periodic far fields that
satisfy de Broglie's dispersion relation. These give rise to wave-like
interference phenomena when particles interact with other matter, thereby
resolving the wave-particle duality paradox. The metron solutions and all
particle interactions on the microphysical scale (with the exception of the
kaon system) satisfy strict time-reversal symmetry, an arrow of time
arising only at the macrophysical level through the introduction of
time-asymmetrical  statistical assumptions. Thus Bell's theorem on the
non-existence of deterministic (hidden variable) theories, which depends
crucially on an arrow-of-time, is not applicable. Similarly, the periodic
de Broglie far fields of the particles do not lead to unstable radiative
damping, the time-asymmetrical outgoing radiation condition being replaced
by the time-symmetrical condition of zero net radiation.

Assuming suitable polarization properties of the metron solutions, it can
be shown that the coupled field equations of the Maxwell-Dirac-Einstein
system as well as the Lagrangian of the Standard Model can be derived to
lowest interaction order from the Einstein vacuum equations. Moreover,
since Einstein's vacuum equations contain no physical constants (apart from
the introduction of units, namely the velocity of light and a similar scale
for the harmonic dimensions, in the definition of the flat background
metric), all physical properties of the elementary particles (mass, charge,
spin) and all universal physical constants (Planck's constant, the
gravitational constant, and the coupling constants of the electroweak and
strong forces) must follow from the properties of the metron solutions. A
preliminary inspection of the structure of the solutions suggests that the
extremely small ratio of gravitational to electromagnetic forces can be
explained as a higher-order nonlinearity of the gravitational forces within
the interior metron core. The gauge symmetries of the Standard Model follow
from geometrical symmetries of the metron solutions. Similarly, the parity
violation of the weak interactions is attributed to a reflexion asymmetry
of the metron solutions (in analogy to molecules with left- and
right-rotational symmetry), rather than to a property of the basic
Lagrangian. The metron model also yields further interaction fields not
contained in the Standard Model, suggesting that the Standard Model
represents only a first-order description of elementary particle
interactions.

While the Einstein vacuum equations reproduce the basic structure of the
fields and lowest-order interactions of quantum field theory, the particle
content of the metron model has no correspondence in quantum field theory.
This leads to an interesting interpretation of atomic spectra in the metron
model. The basic atomic eigenmodes of quantum electrodynamics appear in the
metron model as the scattered fields generated by the interaction of the
orbiting electron with the atomic nucleus. For certain orbits, the
eigenmodes are in resonance with the orbiting electron. In this case, the
eigenmode and orbiting electron represent a stable self-supporting
configuration. For circular orbits, the resonance condition is identical to
the integer-action condition of the Bohr orbital model. Thus the metron
interpretation of atomic spectra yields an interesting amalgam of quantum
electrodynamics and the original Bohr model. However, it remains to be
investigated whether higher-order computations of the metron model are able
to reproduce atomic spectra to the same high degree of agreement with
experiment as QED. On a more fundamental level, the basic questions of  the
existence, structure, stability and discreteness of the postulated metron
solutions still need to be addressed. However, it is encouraging that,
already on the present exploratory level, the basic properties of
elementary particles and fields, including the origins of particle
properties and the physical constants, can be explained within a unified
classical picture based on a straightforward Kaluza-Klein extension to a
higher dimensional space of the simplest vacuum form of Einstein's
gravitational equations.

\newpage

\section{Unification and Beyond}
\label{Unification and Beyond}

Since the conception of quantum theory some seventy years ago, the
development of physics has been beset by a two-fold dichotomy. Despite
intensive efforts, attempts to unify quantum theory with the second basic
pillar of modern physics, general relativity, have so far proven
unsuccessful\footnote{cf. Maiani and Ricci, 1996. Most of these approaches,
such as the current string theories or the Ashtekar (1988) programme,  have
been based on attempts to quantify gravity, rather than to question the
basic premises of quantum theory .}. At the same time, the revolutionary
innovation of quantum theory, the rejection of the concept of
mathematically defined objects at the microphysical level, has divorced
physics from its traditional objectivistic foundations that have continued
to form the basis of advances in all other areas of science\footnote{The
conceptual stagnation in the development of the foundations of physics may
be constrasted, for example, with the spectacular advances made in
microbiology by systematically developing models based on clearly defined
objects, such as DNA molecules, messenger and transfer RNA molecules,
vector viruses, etc.}.

In the following I shall outline an ``objectivistic'' (sometimes termed
``realistic'') ``metron'' theory of fields and particles that I believe is
able to resolve both dichotomies. The basic concept is that one can define
physical objects, in the classical sense, that solve the wave-particle
duality conflict of microphysics by exhibiting both corpuscular and
 wave-like phenomena. The objects represent soliton-type solutions of the
Einstein vacuum equations in eight (or higher) dimensional space. The
extension of Einstein's gravitational equations to a higher dimensional
space is motivated, as in the original approach of Kaluza (1921) and Klein
(1926), by the desire to include further forces in the framework of
Einstein's elegant general relativistic theory of gravity. At least eight
dimensions are needed to include all forces. The reduction of Einstein's
equations to the vacuum form is an important further step that enables the
properties of matter to be derived from the nonlinear structure of the
soliton solutions, rather than postulating the existence of matter and
inserting the relevant source terms into the field equations {\it a
priori}. The soliton solutions
(`` metrons'') are locally concentrated in physical  three-dimensional
space, thereby exhibiting corpuscular-like properties, but at the same time
carry periodic far fields that produce wave-like interference phenomena.
They are also periodic with respect to (or independent of) the extra-space
coordinates.  The metron model yields the Einstein gravitational equations
with matter in physical spacetime, together with the basic field equations
and symmetries of quantum field theory (QFT), as summarized in the Standard
Model of elementary particles.

The motivation for developing a unified theory of fields and particles that
overcomes the conceptual paradoxes of the Copenhagen school is not only
aesthetic. The development of a new view of physics necessarily yields new
insights, leading to new questions with new answers. Thus I shall show that
the  metron model is able to derive or explain the origin of all
fundamental particle properties that are introduced axiomatically in
classical gravity and quantum field theory: mass and electric charge; the
composition of the elementary particle spectrum; particle symmetries and
symmetry breaking; the nature of physical forces, including the exceptional
weakness of the gravitational force; and  all physical constants.  Since
the Einstein vacuum equations contain no free constants (apart from the
defining spatial scales introduced in the normalization of the background
metric), all of these properties must necessarily follow from the nonlinear
structure of the metron solutions.

Whether or not the metron theory will satisfy all these expectations
remains to be investigated. In a first analysis (Hasselmann, 1996a,b,
1997a,b, referred to in the following as I - IV, respectively), it was
shown that appropriate soliton-type solutions of the n-dimensional Einstein
vacuum equations, if they exist, yield the standard Einstein gravitational
equations in four-dimensional spacetime, including the energy-momentum
source terms; the coupled field equations of the Maxwell-Dirac-Einstein
system, including the elementary charge; atomic spectra; the ratio of
gravitational to electromagnetic forces; and the basic structure and
symmetries (with differences in detail) of the remaining coupled field
equations of QFT, in accordance with the Standard Model. Soliton-type
solutions of the posutulated dynamical structure were furthermore computed
explicitly for a simplified scalar Lagrangian that captured the basic
nonlinearities of the Einstein vacuum Lagrangian while suppressing its
tensor complexities. However, numerical computations for the full Einstein
system in
n-dimensional space have still to be carried out. In the following summary,
I shall outline the principal concepts of the metron model, without
entering into the details of the mathematics presented in I - IV.

\section{Is a unified deterministic theory of fields and particles
feasible?}
\label{Is a deterministic}

A program to develop a unified theory of fields and particles based on
classical, objectivisic physics must face two basic objections:

1) As a method of categorizing and computing an enormous variety of
phenomena -- some, as in quantum electrodynamics, at very high accuracy
-- quantum theory has proven immensely successful. It appears intrinsically
unlikely that a theory based on entirely different concepts will be able to
reproduce the detailed results of standard quantum theory. The response to
this objection is that the metron model yields the basic coupled field
equations of QFT to lowest interaction order, and is thus able to reproduce
the principal results of QFT.  However, at higher order the metron model
departs from standard QFT: the interaction terms of the Lagrangians differ
beyond the lowest coupling order, and the metron model has no closed loop
contributions, divergences or renormalization formalism. Thus, it remains
to be seen to which  level of accuracy the two theories are, in fact,
equivalent. It is of interest in this context that Barut (1988) has
claimed that classical tree-level computations of the hydrogen atom
spectrum agree better with measurements  than the computations of quantum
electrodynamics.

Despite the close equivalence of the field content of the two
theories, the metron picture is  fundamentally different from
that of standard quantum field theory. The metron model is characterized
not only by the basic field equations, but also by the discrete corpuscular
solutions of the field equations. Thus, discrete atomic spectra are
represented in the metron model not only by the eigensolutions of the
Schr\"odinger equation (or the corresponding relativistic Maxwell-Dirac
equations), but also by the discrete orbits of the associated electrons
that are resonantly coupled to the eigenmodes. The theory therefore
effectively combines quantum electrodynamics with the original Bohr orbital
model (see Section~\ref{Atomic spectra}).

2) Apart from its practical success, quantum theory is also generally
believed to be the only feasible approach to resolving the wave-particle
duality paradoxes of microphysics. Although von Neumann's original "proof"
that the duality paradoxes cannot be resolved within the framework of
classical objectivistic physics has been demonstrated by Bohm (1952) and
Bell (1964) to rest on invalid assumptions, Bell himself has provided an
alternative, widely accepted proof. In his celebrated theorem on the
Einstein-Podolsky-Rosen experiment, Bell demonstrated that under general,
apparently plausible conditions, it is not possible to explain the paradox
of the EPR experiment in terms of a classical hidden-variable theory.
However, Bell also pointed out that his proof rested critically on the
assumption of causality, in the sense of the existence of an arrow of time.
As discussed below, the metron model is based on strict time-reversal
symmetry on the microphysical level, so that Bell's theorem does not apply.
An interpretation of the EPR experiment from the viewpoint of time-reversal
symmetry is given in III.

Time-reversal symmetry on the microphysical level is also the reason that
the metron solutions are able to support stable periodic far fields. In
contrast to the standard view that such solutions must decay through
radiation to infinity, the time-asymmetrical boundary condition of outgoing
radiation is replaced in the metron model by the time-symmetrical condition
of zero net radiation (equal and opposite ingoing and outgoing radiation).
This permits stable standing-wave solutions. The outgoing-radiation
condition arises only later at the macrophysical level through the
introduction of time-asymmetrical statistical assumptions (see discussion
in Einstein, 1909, Wheeler and Feynman, 1949,  III and also Gutzwiller,
1990).

\section{Wave-particle duality}
\label{Wave-particle}

Quantum theory was invented to resolve the wave-particle paradox. The
difference between the quantum-theoretical and metron approach to this
problem is best illustrated by an experiment in which a uniform stream of
particles of given momentum is observed to exhibit  wave-like interference
phenomena on interacting with an object. Instead of the traditional single-
or double-slit experiment, I consider the case of Bragg scattering at a
periodic lattice, as this is more amenable to elementary analysis.

If the particle stream is sufficiently weak, it is possible to measure
individual particles immediately after they leave the particle source, and
again after they have been Bragg scattered by the lattice, when they
impinge, for example, on a particular counter of a particle counter array
located behind the lattice. In the naive classical view, it is clear that
one is
measuring individual particles, with reasonably well defined initial and
final positions and momenta\footnote{That these underly the Heisenberg
uncertainty relation is not relevant for the present discussion}. However,
in apparent conflict with this result, the statistical distribution of
particle counts is found to correspond to the interference pattern of a
periodic wave incident on the lattice.

The quantum theoretical response to this finding is that one is faced with
an insurmountable contradiction that can be circumvented only by rejecting
both premises that one is observing either a particle or a wave. One
introduces instead a general formalism for predicting the statistical
outcome of microphysical experiments without defining what microphysical
objects actually are. In fact, it is stated that physical objects in the
classical sense do not exist at the microphysical level. In contrast to the
prevalent Copenhagen view, the metron concept is based on the alternative,
rather obvious conclusion that one is indeed observing real, discrete
particles, but that the interactions of the particles with other objects
exhibit
wave-like interference phenomena\footnote{This is reminiscent of another
Copenhagen school of thought, expounded by Hans Christian Anderson in his
tale of the emperor's new clothes.}.

To produce interference phenomena, the particles must carry periodic far
fields. In the usual classical picture, this would lead to radiative
damping, and would thus not be acceptable for a stable particle. However,
as mentioned above, at the microphysical level of fundamental particles,
radiative damping is excluded in the metron model as a time-asymmetrical
process. Radiative damping is explained within the framework of
time-symmetrical elementary interactions as a macrophysical phenomenon
involving irreversible interactions of an accelerated  test particle  with
a non-time-symmetrical statistical ensemble of distant particles  (Wheeler
and Feynman, 1948, III, see also Einstein, 1909).

In order to reproduce the interference characteristics of de Broglie waves,
the wavenumbers of the particles' periodic far fields must satisfy de
Broglie's relation\footnote{De Broglie (1956) and Bohm (1952) have
similarly proposed a theory of real particles and de Broglie waves.
However, in contrast to the de Broglie-Bohm pilot wave model, the de
Broglie fields in the present case are not regarded as  separate entities
guiding the particles, but rather as integral components of the particles'
fields. The relation between the de Broglie-Bohm theory and the metron
model will be discussed in a separate paper.}
\begin{equation} \label{w1}
k_{\mu} = p_{\mu}/\hbar = m u_{\mu}/\hbar,
\end{equation}
where $p_{\mu}$ is the four-momentum,  $u_{\mu}$ the velocity and $m$ the
rest mass of the particle. In the particle restframe, the field is periodic
in time with (angular) frequency
\begin{equation} \label{w1a}
 \omega = m c^2/\hbar,
\end{equation}
and has infinite wavelength (in the following natural units will be used,
with $c=\hbar=1$).

The interaction of the periodic far field of the particle (denoted in the
following simply as the particle's de Broglie field) with the periodic
lattice generates a set of scattered waves with wavenumbers
$k^{(s)}_{\mu}$ given by the Bragg relation
\begin{equation} \label{w2}
k^{(s)}_{\mu} = k^{(i)}_{\mu} +  k^{(l)}_{\mu},
\end{equation}
where $k^{(i)}_{\mu}$ is the wavenumber of the incident particle and
$k^{(l)}_{\mu}$ is one of the (not necessarily fundamental) periodicity
wavenumbers of the lattice. In order to be able to propagate, the scattered
waves must satisy the de Broglie free-wave  dispersion relation (see below)
\begin{equation} \label{w3}
k^{(s)}_{\mu} k^{(s) \mu} = -  \omega^2,
\end{equation}
where the  background spacetime metric is defined as $\eta_{\nu \mu} = $
diag$ (1,1,1,-1)$.

The relations (\ref{w2}) and (\ref{w3}) are identical to the usual
scattering computations of quantum theory, which determine the resonant
Bragg directions of the scattered particle beams. However, in the present
deterministic particle picture, the scattered field cannot be simply
identified with the particle beam, and one must ask further: how does the
scattered Bragg  field affect the individual particle trajectories?

Consider a particle that is deflected by the lattice into a scattered
velocity $u^{(s)}_{\lambda}$. In the particle restframe, it will experience
the scattered de Broglie field with the ``frequency of encounter''
\begin{equation} \label{w4}
\omega^{(e)} = - k^{(s)}_{\lambda} u^{(s) \lambda}.
\end{equation}
If $\omega^{(e)}$ differs from the intrinsic particle frequency $ \omega$
associated with its de Broglie wave, the interaction of the particle with
the periodic scattered wave will average to zero: the scattered field has
no impact on the particle trajectory. However, if $\omega^e =  \omega$, the
scattered field is in resonance with the intrinsic periodicity of the
particle, and the quadratic interaction of the scattered field with the
particle's intrinsic de Broglie field yields a mean force. It can be
readily verified (III) that this second  particle resonance condition
requires a scattered particle velocity  in the direction of the Bragg
scattered wave,
\begin{equation} \label{w5}
u^{(s)}_{\lambda} =  \omega^{-1} k^{(s)}_{\lambda}.
\end{equation}

Expressed in terms of the interaction Lagrangian of the scattered particle,
the particle resonance with its scattered wave produces
$\delta$-function-type  canyons in the potential energy, which effectively
trap the particles in the preferred Bragg scattering directions. Thus the
deterministic particle picture yields qualitatively similar results to the
wave scattering computations of quantum theory: the scattered particle
beams are concentrated in the discrete resonant Bragg scattering
directions. A more detailed presentation of the trapping mechanism is given
in III.

This simple example was described in some detail as it illustrates an
important feature of the metron model: the model contains essentially the
same field content as standard quantum field theory, but goes beyond QFT in
providing also predictions for the trajectories of objectively existing
discrete particles. The metron computations of the field interactions
follow  closely the quantum theoretical analysis based on resonant
wave-wave interactions\footnote{The resonance conditions express the
conservation of four-momentum. A similar formalism can be applied also to
resonant wave-wave interactions between geophysical wave fields, see
Hasselmann (1966).}, while the computations of the particle trajectories
require the consideration of further wave-particle interactions. In the
present case, the dominant wave-particle interactions were  determined by a
second resonance condition. A similar wave-particle resonant-interaction
condition is found to explain the origin of discrete electron orbits in the
metron model of atomic spectra (Section~\ref{Atomic spectra}). A more
detailed discussion of the relation between the metron model and the
standard QFT picture of elementary particles and fields, including a
discussion of the Heisenberg uncertainty principle\footnote{Although the
position and momentum of a particle can be simultaneously defined in
principle in the metron picture, Heisenberg's uncertainty relation follows
in practice from the impossibility of reducing an initial uncertainty
$\Delta x.\Delta p > \hbar$ in the position and momentum of a particle
through a measurement process involving interactions with another system
that underlies a similar uncertainty -- see Bohm (1952).}, boson-fermion
statistics\footnote{See also next section} and other fundamental features
of QFT, is given in I.

\section{Basic equations and structure of the metron model}
\label{Basic equations}

Having recognized that the wave-particle duality paradox can be overcome if
one accepts the notion that particles carry periodic de Broglie far fields,
how can one introduce particles with this property? One approach could be
to simply postulate the existence of discrete particles with the required
properties, together with the fields needed to describe their interactions.
This would be analogous to the axiomatic introduction of the various
particle  fields and their interactions in QFT. However, if one regards
this as unsatisfactory, one must explain the existence and properties of
particles as the solutions of some set of (nonlinear) field equations.
Assuming that the ultimate goal is to develop a unified theory, the
equations  must contain as a minimum both Einstein's gravitational
equations and Maxwell's equations. It has been shown by Kaluza (1921) and
Klein (1926) that both systems of equations can be derived from Einstein's
equations in a higher dimensional space. So why not try Einstein's
equations in some
n-dimensional space?

If one wishes to pursue further the goal of explaining the origin of mass,
charge and the coupling constants of the weak and strong interactions, the
source terms in the n-dimensional Einstein equations, in which these
constants would appear, must be dropped.  Thus I take as the basic
equations of the metron model simply the Einstein vacuum equations
\begin{equation} \label{w6}
R_{LM} = 0
\end{equation}
in some n-dimensional space, where $g_{LM}$ denotes the metric, $R_{LM}$
the Ricci curvature tensor,
\begin{equation} \label{3.21a}
R_{LM} := \partial_{M} \Gamma^{N}_{L N}
-  \partial_{N}\Gamma^{N}_{LM}
+ P_{LM},
\end{equation}
\begin{equation} \label{3.22}
P_{L M} := \Gamma^{N}_{L O}\,
\Gamma^{O}_{M N} -
\Gamma^{N}_{LM}\, \Gamma^{O}_{N O}
\end{equation}
and the connection (Christoffel symbol) is given by
\begin{equation} \label{3.23}
\Gamma^{L}_{M N}  : =  \frac{1}{2} \,  g^{L
O}   \,
\left[
\partial_{M}   g_{O N}   +
\partial_{N} g_{O M} - \partial_{O} g_{M N}
\right].
\end{equation}

I assume that soliton-type solutions of these equations that can be
identified with discrete particles exist. Solitons have been extensively
studied as solutions of the nonlinear equations of fluid dynamics. They
normally represent  a balance between  linear dispersion, which tends  to
smooth out disturbances, and nonlinearities, which tend to concentrate
disturbances (often towards a discontinuous shock). Solutions are normally
found only in two dimensions, since in three dimensions, linear geometrical
dispersion dominates over the nonlinearities.

\begin{figure}[!t] \centering
$$\epsfig{file=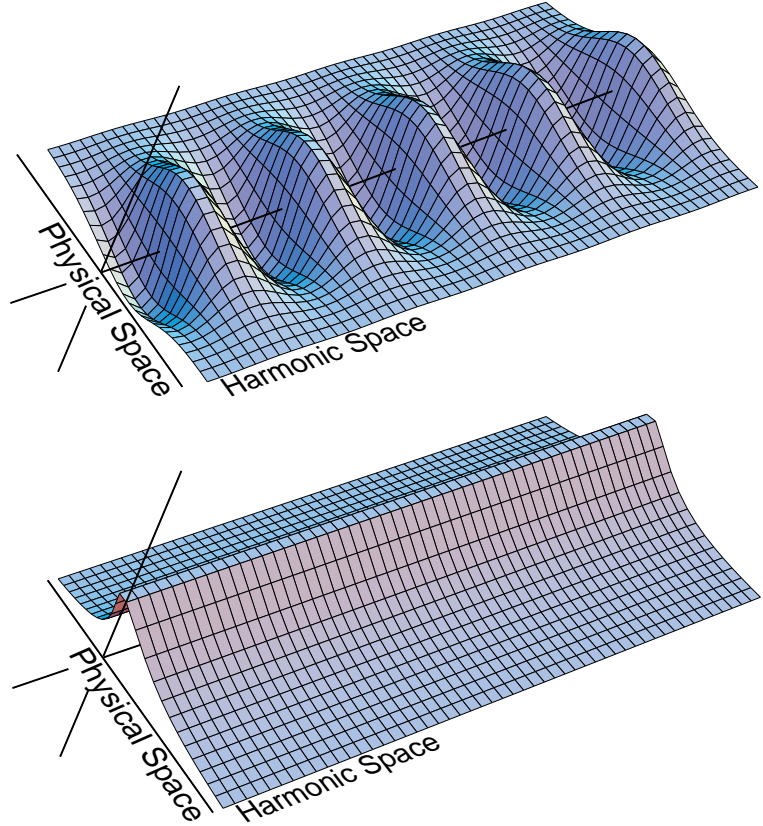, width=10cm}$$
\caption{Schematic diagram of trapped-mode (upper panel) and
wave-guide (lower panel)
components of a metron particle. Plotted in the lower panel is the inverse
of the phase speed; the reduced phase speed in the core region represents a
wave-guide that traps the wave through total reflection near the edge of
the core region (from I).}
\label{figwolf1}
\end{figure}

The postulated metron solutions of the Einstein vacuum equations are
soliton solutions of a basically different nature. They are comprised of a
nonlinear superposition of a number of ``partons''. Each parton is
localized in three-dimensional physical space and periodic with respect to
time and extra-space (termed ``harmonic space'' in the following). The
periodic wave fields propagate in a mean metric field that is distorted in
three-dimensional space and thereby acts as a wave-guide: the wave phase
velocities are reduced within the central core region of the particle,
resulting in the  total reflection of the waves near the edge of the
particle core. The distortion of the metric, in turn, is produced by the
nonlinear (to lowest order, quadratic ``radiation stress'' or ``current'')
self- and cross-interactions of the wave field components. Since the wave
amplitudes are largest within the metron core, this is also the region of
the largest metric distortions. Thus the metron solutions represent a
mutually supporting  coupled trapped-mode/wave-guide system
(Figure~\ref{figwolf1}).  Similar mutually sustained trapping interactions
between waves and the medium in which they propagate have been studied in
various geophysical applications (e.g. Garrett, 1976, Haines and
Malanotte-Rizzoli, 1991). The existence of soliton-type solutions based on
this trapped-mode/wave-guide interaction mechanism has been demonstrated in
the present context in I for a simple scalar Lagrangian obtained by
projecting the Einstein tensor equations onto a few modes. Computations of
metron solutions of the full system of Einstein vacuum equations are
currently in progress.

Since the individual parton components of the metron solutions are periodic
in harmonic space with a small periodicity scale $1/k$ (the wavenumber $k$
is found to be proportional to the coupling strength, see below), the
n-dimensional space is compactified in the metron solution into a thin
(n-4)-dimensional sheet with non-trivial structure only in the remaining
4-dimensional physical spacetime.

In the following I assume that trapped-mode soliton-type solutions of these
equations exist. Thus the metric can be represented as  a superposition of
parton fields $(p)$ that are periodic in harmonic space and time and
localized in physical space,
\begin{equation} \label{w2.1}
g_{LM} = \eta_{LM}   + \sum_{p} g_{L
M}^{(p)}
\end{equation}
where $g_{LM}$ is the total metric field, the background metric is given by
$\eta_{LM} =$ diag $(1,1,1,-1, 1,\ldots,1)$\footnote{In II and IV, a
non-Euclidean background harmonic metric, $\eta_{LM} =$ diag $(1,1,1,-1,
\pm 1,\ldots,\pm 1)$, is also considered.}, and the periodic  parton
components
\begin{equation} \label{w2.2}
 g_{LM}^{(p)} := \hat g_{LM}^{(p)}(x) \exp
(iS^p)
+ \mbox{compl. conj.},
\end{equation}
have phase functions
\begin{equation} \label{w2.3}
S^{p} : = k^{(p)}_{A} x^{A}
\end{equation}
with constant harmonic (extra-space) wavenumber vectors $k^{(p)}_{A}$  and
amplitudes $\hat
g_{LM}^{(p)}(x)$ that are functions of physical
spacetime
$x$
only. In the parton restframe, the amplitudes are localized in physical
space ${\bf x}$ and periodic in time $t=x^4$,
\begin{equation} \label{3.3}
\hat g^{(p)}_{LM}(x) = \tilde g^{(p)}_{LM}({\bf x})
\exp \left( - i
\omega^{(p)}t \right ),
\end{equation}
where the frequency $ \omega^{(p)} = -k_4^{(p)}$ is identified with the
parton mass. There exist also massless partons with  $ \omega^{(p)}=0$. The
mean ``wave-guide'' metric-field component is included in the
representation (\ref{w2.1}) as a component with zero  frequency and
harmonic wavenumber.

 The index and coordinate notation used here and in the
following is
defined in Table \ref{ta2.1}.  Non-tensor indices,
which are excluded from the
summation convention, are placed in parentheses.
\begin{table}
\begin{tabular}{|lcl|}
\hline
\multicolumn{1}{|c}{space}  &  components  &
\multicolumn{1}{c|}{vector}\\
\hline
full n-dimensional space     & $x^{L}$        & $X =
(x^1 ,x^2
,\cdots, x^n )$\\
three-dimensional physical space & $x^i$    & ${\bf x} =
(x^1 ,x^2 ,x^3)$\\
four-dimensional physical spacetime & $x^{\lambda}$& $ x  =
(x^1
,x^2 ,x^3 ,x^4)$\\
$(n - 4)$-dimensional harmonic space & $x^A$ &
x = $(x^5 ,x^6 , \cdots,x^n)$ \\
\hline
\end{tabular}
\caption {Index and coordinate notation}
\label{ta2.1}
\end{table}

The parton components interact at various orders of nonlinearity and
consist generally of the fundamental and higher-harmonic components of a
basis set of partons (normally fermions, for example, the three quarks of
the proton/neutron system).

The representation (\ref{w2.1}) of the full metric field $g_{LM}$ as a sum
of perturbations $g_{LM}^{(p)}$ superimposed on the flat-space background
metric $\eta_{LM}$ implies that interactions between the perturbations can
be represented as a nonlinear series expansion (which need not be
restricted to weak interactions, however). Furthermore, through the
assumption of a flat background metric we disregard here the problem of the
imbedding of the local background  metric in a more general curved space on
cosmological scales.

The coordinate system can be chosen such that all parton components satisfy
the gauge condition
\begin{equation} \label{2.7}
\partial^{L} h_{LM}^{(p)} = 0,
\end{equation}
where
\begin{equation} \label{2.8}
h_{LM}^{(p)} : = g_{LM}^{(p)} -
\frac{1}{2} \eta_{LM} \,
g_{N}^{(p)N}
\end{equation}
is the trace-reversed representation of the metric perturbation.

In the linearized approximation, the gravitational equations reduce in this
case for each parton component to the n-dimensional wave equation
\begin{equation} \label{2.4}
\partial _{N} \partial ^{N} g_{LM}^{(p)}
= 0,
\end{equation}
which yields for the parton amplitudes the Klein-Gordon
equation
\begin{equation} \label{2.5}
\left( \Box -  \hat \omega^{(p)2} \right) \hat
g^{(p)}_{LM} = 0,
\end{equation}
where the  ``harmonic'' frequency $\hat \omega^{(p)}$ is defined
as\footnote{For linear fields, the harmonic and de Broglie frequencies are
the same, $\hat \omega^{(p)}=  \omega^{(p)}$, but in the general nonlinear
case, $\hat \omega^{(p)} \neq  \omega^{(p)}$.}

\begin{equation} \label{2.6}
 \hat \omega^{(p)} : = \left( k_{A}^{(p)} k^{A}_{(p)} \right)^{1/2}.
\end{equation}

The  field equations of QFT  and classical gravity, including the source
terms,  are obtained from Einstein's  vacuum equations in n-dimensional
space by assigning each parton component $g_{LM}^{(p)}$ to one of the basic
fields $\phi^{(a)}$ of QFT or classical gravity. The relation between the
components of the n-dimensional metric and  the standard fields of QFT and
gravity is established through a set of  polarization matrices
$P_{(a)LM}^{(p)}$,
\begin{equation} \label{3.10}
\hat g^{(p)}_{LM} = \sum_{(a)} P^{(p)}_{(a)LM} \phi^{(a)}.
\end{equation}
The assignment of the different QFT and classical gravity fields to the
various components of the n-dimensional metric is summarized in
table~\ref{ta3.1}. Fermions (and the Higgs field) are assigned to
harmonic-sector metric components, while bosons are represented by  metric
components with mixed spacetime-harmonic indices. The basic trapped-mode
fields of the metron solutions are comprised of fermions, the bosons being
auxiliary fields generated by quadratic (difference) interactions between
fermions. The corpuscular properties of matter (in particular, the particle
mass) are determined by the fermion fields. Bosons (for example, photons)
are not regarded as particles in the metron model, but as classical fields.
They derive their corpuscular-like properties from the transitions between
discrete particle states that they mediate (Lorentz, 1904). The Fermi-Dirac
and Einstein-Bose statistics of fermions and bosons (in particular, the
Pauli exclusion principle) follows in the metron model simply from the
observation that it is not possible for two existing particles (fermions)
to be at the same position at the same time, while it is quite feasible to
superimpose the associated boson fields of different particles.
\begin{table}
\begin{tabular}{l||c|c}
& spacetime indices & harmonic indices\\
~&~&~ \\
 \hline
 \hline
~&~&~ \\spacetime indices & gravity $g_{\lambda \mu}$ & bosons
$ \hat g^{(b)}_{\lambda A} = B_{\lambda} a^{(b)}_A $ \\
~&~&~ \\
 \hline
~&~&~ \\
harmonic indices & bosons $ \hat g^{(b)}_{A \lambda}= a^{(b)}_A
B_{\lambda} $ & fermions $\hat g^{(f)}_{AB}= P^{(f)a}_{AB}
\psi_a $ \\
                 && scalars  $\hat g^{(s)}_{AB}= P^{(s)}_{AB}
\varphi $\\
\end{tabular}
\caption[ix] {Associated metric tensors and polarization relations for
gravitational fields $g_{\lambda \mu}$,
vector bosons $B_{\lambda}$,
fermions $\psi_a $ and
scalar fields $\varphi$ ($a^{(b)}_A=$ const = bosonic polarization factor).
Metric fields represent deviations from the flat-space background metric
$\eta_{LM}$ and represent, in general, the complex amplitudes of periodic
fields, cf. eqs.~(\protect{\ref{w2.2}}), (\protect{\ref{3.10}}).}
\label{ta3.1}
\end{table}

For finite-mass fields, which are periodic in harmonic space, the
amplitudes $\phi^{(a)}$ and polarization matrices $P^{(p)}_{(a)LM}$ are
complex. This enables the representation of complex fields such as Dirac
four-spinors in terms of the tensor components of the real n-dimensional
metric\footnote{It may appear surprising at first sight that  half-integer
spin fields can be represented in terms of an
integer-spin metric field. However, it should be noted that the
polarization relations for Dirac fields apply to the harmonic rather than
the spacetime sector of the n-dimensional metric, see Table~\ref{ta3.1} and
the discussion in II.}.

As pointed out, since the Einstein vacuum equations (\ref{w6}) contain no
physical constants beyond the scale normalization, all particle constants
and the universal physical constants that describe their interactions must
be determined in the metron model by the metron solutions. In addition to
the normal invariance with respect to coordinate transformations
(diffeomorphisms), the Einstein vacuum equations are invariant under an
arbitrary common scale change in all coordinates (without otherwise
modifying the metric field).  One can therefore arbitrarily assign a unit
to one of the fundamental harmonic space wavenumbers of the metron
solutions. The values of all other wavenumbers with respect to this
reference wavenumber, as well as the magnitudes and polarization structures
of the metric field components, are then determined by the properties of
the metron solutions.

In discussing the relation of the metron model to QFT and classical
gravity, one must consider both the field and  particle content of the
metron model.  The field content can be directly related to QFT and to the
field equations of classical gravity (with the exclusion of the
energy-momentum source terms). However, the metron particle picture has no
direct counterpart in QFT and can therefore be related only to  the
classical picture of interacting point particles (see, for example, Wheeler
and Feyman, 1949, for the case of electromagnetic interactions). It can be
shown that the particle picture of the metron model reproduces the
point-particle source terms of both Maxwell's equations and classical
gravity, at the same time determining the relevant particle properties and
physical coupling constants. I discuss first field-field interactions and
subsequently field-particle (or, equivalently,  particle-particle)
interactions.

\subsection{Field interactions}
\label{Nonlinear}
To derive the field equations of QFT and classical gravity from the metron
field equations, one simply substitutes the postulated polarization
relations (\ref{3.10}) into the Einstein vacuum equations and separates the
resulting linear and nonlinear terms with respect to the basic parton
fields. It is rather surprising that one does indeed recover in this manner
all the basic field equations, with the correct lowest-order interaction
terms, of QFT and classical gravity, including not only the
Maxwell-Dirac-Einstein system (II), but also the weak and strong
interactions, with associated fermions, bosons, and even the Higgs field,
as summarized in the Standard Model (although with minor differences in
detail, see IV).

The relations between the various components of the Ricci tensor and the
field equations for the Maxwell-Dirac-Einstein system are indicated in
Table~\ref{tawolf3}. The field equations of classical gravity (excluding
the source terms, that are discussed in the next subsection) follow
trivially by identifying the physical spacetime sector $g_{\lambda \mu}$ of
the n-dimensional metric with the classical gravitational metric. The
harmonic-space subsegment $g_{AB}$ of the n-dimensional metric is
identified with fermion fields (in the general case of the Standard Model,
also the Higgs field). To recover the four-spinor properties of the fermion
fields, the dimension of full space must be at least eight. A particularly
simple form of the spinor polarization matrices follows for a
four-dimensional  harmonic sub-space with Euclidean background metric,
$\eta_{AB} = $ diag $(1,1,1,1)$, but other models are conceivable.

The electromagnetic field is represented by the metric component $g_{5
\lambda}$. In the general case of the Standard Model, bosons are
represented by mixed-index metric fields $g_{\lambda A}$. They are
generated by quadratic difference interactions between fermions. The
quadratic sum interactions and higher order interactions have no
counterpart in the Standard Model, which appears therefore from the metron
viewpoint only as an approximation.

Linear terms are shown on the left hand sides of Table~\ref{tawolf3}; they
are obtained by direct substitution of the polarization forms (\ref{3.10})
(see also table~\ref{ta3.1}) into the linearized n-dimensional vacuum
gravitational equations and involve no cross-coupling with other sectors of
the gravitational metric. In contrast, the nonlinear source terms on the
right-hand sides arise through cross-coupling to other sectors. The matter
source term (the energy-momentum tensor $T_{\lambda \mu}$) of the
4-dimensional gravitational equations is determined by higher-order
interactions than the source term (the electromagnetic current) of the
Maxwell equations (see discussion in the next sub-section).
\begin{table}
\begin{tabular}{l||c|c}
& 4d-spacetime & harmonic sector\\
~&~&~ \\
 \hline
 \hline
~&~&~ \\
4d-spacetime & 4d-gravity: $R_{\lambda \mu}=0 $ & Maxwell:
$R_{5 \lambda}=0$
 \\
~&$\rightarrow \Box\;h_{\lambda \mu} = - 2\,G\,T_{\lambda \mu}$
& $\rightarrow \Box\;A_{\lambda} = q (\psi \gamma_{\lambda} \psi)$ \\
~&~&~ \\
 \hline
~&~&~ \\
harmonic sector & Maxwell: $R_{\lambda 5}=0$ & Dirac: $R_{AB}=0 $ \\
~& $\rightarrow  \Box\;A_{\lambda} = q (\psi \gamma_{\lambda} \psi)$
& $\rightarrow \left(\gamma^{\lambda} \partial_{\lambda} + \omega \right)
\psi = i q A_{\lambda} \psi $\\
\end{tabular}
\caption[ix] {Relation between the Ricci components $R_{LM}$ of the
n-dimensional metric and the field equations of the Maxwell-Dirac-Einstein
system. $A_{\lambda}, q, \psi, h_{\lambda \mu}, G$ and $T_{\lambda \mu}$
denote the electromagnetic potential, charge, Dirac 4-spinor,
trace-reversed 4-dimensional gravitational field, gravitational constant
and energy-momentum tensor, respectively (see Section~\protect{\ref{The
Standard Model}} for extension to the Standard Model).}
\label{tawolf3}
\end{table}

The QFT coupling constants follow from the polarization matrices and are
proportional generally to the wavenumbers of the interacting parton
components. The gauge symmetries of the Standard Model are explained as
geometrical symmetries of the metron solutions. Thus the parity violation
of the weak interactions, for example, is attributed to a geometrical
property of the metron solutions themselves (in analogy with the existence
of molecules with left- or right-rotational symmetry) rather than
\nopagebreak to a reflectional asymmetry of the basic Lagrangian.

\subsection{Field-particle interactions}
\label{Field}
For QFT, it is sufficient to specify the  coupling between the basic fields
of the theory. There exists then a formalism for computing (in principle)
the statistical outcome of the interactions between particles of any given
configuration. In the metron model, however, the field interactions are
only part of the picture. All fields are rooted in real discrete particles,
and -- as discussed in the example of Bragg scattering -- their
interactions are of interest mainly through their effect on the
trajectories (or stabilities) of the associated particles. Since the
concept of an objective, localized particle is foreign to QFT, the metron
picture of interacting particles must be related to the classical picture
of interacting point particles. This is restricted,  however, to the
gravitational and electromagnetic distant-interaction fields, for which the
interacting particles can indeed be treated as point particles
characterized by a mass and charge, with a universal gravitational coupling
constant $G$ (the electromagnetic coupling constant is absorbed in the
definition of charge).

The classical gravitational and electromagnetic coupling between a set of
point particles $(i)$ can be expressed (II) by the action
\begin{equation} \label{4.6}
W_{cl} : = W_g + W_A + W_{pg} + W_{pA},
\end{equation}
consisting of the sum of two field action integrals
\begin{eqnarray} \label{4.7}
W_g     &: =& -  \frac{1}{4 G}\,   \int
              \left\{
                \partial_{\lambda} g^{\mu\nu}
                \partial^{\lambda}  g_{\mu  \nu}  -
\frac{1}{2}  \,
                \partial_{\lambda} g^{\mu}_{\mu}
                \partial^{\lambda} g^{\nu}_{\nu}
              \right \}\; d^4x \\
\label{4.8}
W_A    &: =& - \frac{1}{2}\, \int \, F_{\lambda
\mu}F^{\lambda\mu}\; d^4x
\end{eqnarray}
and two particle-path action integrals
\begin{eqnarray}
 \label{4.9}
W_{pg} &: =&  - 2 \,  \sum_i \, \int_{T_{(i)}}\; m_{(i)}
             \left\{
                - g_{\lambda\mu} u^{\lambda}_{(i)} \,
u^{\mu}_{(i)}
             \right \}^{1/2} \; ds \\
\label{4.10}
W_{pA}  &: =&  2   \,      \sum_i   \int_{T_{(i)}}
q_{(i)}A_{\lambda}
             \, u^{\lambda}_{(i)} \; ds,
\end{eqnarray}
where
\begin{equation} \label{w7}
F_{\lambda \mu} = \partial_{\lambda} A_{\mu} - \partial_{\mu} A_{\lambda}
\end{equation}
is the electromagnetic field, $A_{\lambda}$ the electromagnetic potential,
and $m_{(i)}, q_{(i) },T_{(i)}$ the mass, electromagnetic charge and
trajectory, respectively, of particle $(i)$.

Variation of the path integrals with respect to the particle path $T_{(i)}$
yields the particle equations of motion (dropping the particle index
$(i)$)\footnote{The following representation of particle interactions in
terms of mediating fields can be formulated alternatively as direct
particle-particle interactions by expressing the fields as Green-function
integrals over the paths of the particles that generate the fields, cf.
Wheeler and Feynman (1949).}
\begin{equation} \label{4.1}
\frac{du}{ds}^{\lambda}   +    \Gamma^{\lambda}_{\mu
\nu}u^{\mu}u^{\nu}    =
\frac{q}{m}\,F^{\lambda}_{\mu}\,u^{\mu},
\end{equation}
while variation of the action integrals  with respect to the gravitational
field $g_{\lambda \mu}$ and electromagnetic potential $A^{\lambda}$ yields
the field equations.

For the gravitational field one obtains  (linearizing the left hand side,
since we are concerned here only with the interactions with the source
terms)
\begin{equation} \label{4.2}
\Box\;h_{\lambda \mu} = - 2\,G\,T_{\lambda \mu},
\end{equation}
where
$h_{\lambda\mu}
=
g'_{\lambda\mu} - \frac{1}{2} \,\eta_{\lambda \mu}
g'^{\nu}_{\nu}$
is the  trace-reversed representation of the perturbation $g'_{\lambda\mu}
= g_{\lambda \mu} - \eta_{\lambda
\mu}$ of the 4-dimensional gravitational field about the reference
background metric $\eta_{\lambda\mu}$, and the source term is given by the
energy-momentum tensor
\begin{equation} \label{4.4}
T_{\lambda  \mu}  : =  \sum_i  \,m_{(i)}\,
\int_{T_{(i)}}\;ds
\,u_{\lambda}\,
u_{\mu} \delta^{(4)} (x- \xi (s)).
\end{equation}

The corresponding electromagnetic field equations are
\begin{equation} \label{4.3}
\Box \, A^{\lambda} = j^{\lambda},
\end{equation}
where the source term is given by the electric current
\begin{equation} \label{4.5}
j^{\lambda} : = \sum_i \; \int_{T_{(i)}}\;ds \,q^{(i)}
u^{\lambda} \delta^{(4)}
(x- \xi (s)).
\end{equation}

How can one recover these classical field-particle interaction relations
from the metron model? The starting point is the  n-dimensional
gravitational action (II)
\begin{equation} \label{4.13}
W_n : = \int \left|g_n \right| ^{1/2} R \, d^nX,
\end{equation}
where $R := R_L^L$  is the scalar curvature and $\left|g_n \right|$ the
determinant of the metric. Since all fields are assumed to be periodic in
or independent of harmonic space, the integral over harmonic space in
(\ref{4.13}) yields only a normalization factor and can be ignored. Thus
the scalar curvature can be regarded as an harmonic-space average, and the
integral in (\ref{4.13}) restricted to physical spacetime.

\begin{figure}[!t] \centering
$$\epsfig{file=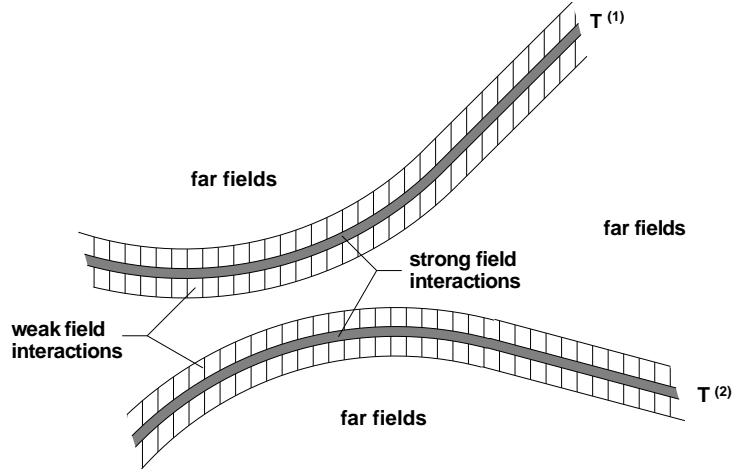, width=10cm}$$
\caption{Metron paths, indicating the core regions of strong field
interactions, the regions of weak field interactions and the linear
far-field regions, that together determine the distant interactions between
quasi-point particles.}
\label{figwolf2}
\end{figure}

By distinguishing between the near- and far-field regions of the integral,
it can be decomposed into a spacetime integral and particle-path integrals,
in accordance with the separate spacetime and path-integral components of
the classical action expression (\ref{4.6}). Formally, the physical
spacetime integral is subdivided into  near-field spacetime tubes
encompassing the metron particle trajectories and the remaining far-field
region (Fig.\ref{figwolf2}). The particle tubes consist of an inner core of
strong interactions that determine the structure of the  trapped
wave-mode/wave-guide metron solution, and an outer mantle in which the
interactions (mainly ``electroweak'') are sufficiently weak to  be treated
as perturbations.  In the far-field regions outside the trajectory tubes,
the  field interactions are neglible.

Substitution of the metron polarization relations (\ref{3.10}) for the
classical gravitational and electromagnetic fields into the gravitational
action (\ref{4.13}) yields in the far-field region the classical action
expressions $W_g$ (eq.(\ref{4.7})) and $W_A$ (eq.(\ref{4.8})), while the
action integrals over the particle trajectory tubes yield the classical
path integral action expressions $W_{pg}$ (eq.(\ref{4.9})) and $W_{pA}$
(eq.(\ref{4.10})). To reduce the spacetime action integrals over the metron
particle-trajectory tubes to  path integrals, the n-dimensional scalar
curvature must be integrated across the tube cross-sections. The
cross-section averages yield the particle mass $m$, charge $q$ and the
gravitational coupling constant $G$ as functions of the metron solution
within the nonlinear near-field region.

Variation of the n-dimensional gravitational action also yields generally
the Einstein-Maxwell-Dirac interaction equations in the outer mantle region
of weak interactions and the complete system of interacting field equations
in (approximate) accordance with the Standard Model within  the  metron
core -- as discussed already in the previous subsection. However, the focus
here is not on the form of the field interactions or the structure of the
metron solutions within the nonlinear tube region, but rather the
integrated net  coupling of the fields in the nonlinear tube region with
the far fields of other particles. It is rather surprising that one does
indeed recover from the metron model the detailed structure of the complete
set of action integrals (\ref{4.7}) - (\ref{4.10}) that describe the
classical gravitational and electromagnetic distant interactions between
quasi-point particles - at the same time determining also the basic
particle and coupling constants.

The gravitational constant $G$ is of special interest. In contrast to the
electromagnetic coupling, which is determined by the lowest-order
interactions of the electromagnetic far field of particle $(i)$ with the
electromagnetic fields of particle $(j)$ in the weak-field-interaction
(mantle) region of particle $(j)$, the gravitational coupling is found to
vanish to this lowest interaction order. It is determined by higher-order
interactions within the metron core.  This explains the extreme weakness of
the gravitational forces (and also the fact that the gravitational coupling
appears only with one sign, i.e. that gravitational forces, in contrast to
electromagnetic forces, are always attractive).

The particle mass appears in the metron formalism through the time
derivative terms of the Ricci tensor.  These yield expressions proportional
to the metron frequency, which must be translated into the metron mass via
Planck's constant (eq. (\ref{w1a})).  Thus in the process of reproducing
the classical distant-interaction relations  for point particles, the
metron model yields not only the particle mass, electric charge and the
ratio of the gravitational to electromagnetic coupling, but also Planck's
constant.

If this appears rather magical, it must be recalled that, since the
Einstein vacuum equations contain no universal constants, all physical
constants derived from the metron model must necessarily follow from the
internal structure of the metron solutions, which contain only one free
scale parameter, a reference wavenumber. Nevertheless, at the level of
analysis outlined here, I have shown only that the existence of general
trapped-mode metron solutions is plausible (I) and have otherwise simply
postulated the structure of the polarization relations (\ref{3.10})
required to reproduce the linearized field equations of classical gravity
and QFT. The main result of the  analysis summarized in Sections
\ref{Nonlinear}, \ref{Field} is that, under this premise, one recovers from
the n-dimensional Einstein vacuum equations not only the linear field
equations, but also the basic structure of the interactions between these
fields. But there remains still the basic challenge of  computing the
metron solutions themselves and demonstrating that stable coupled
wave-mode/wave-guide solutions with the assumed polarization relations do
indeed exist.

\section{Atomic spectra}
\label{Atomic spectra}

A critical test of the metron model is whether it is able to match the
impressive performance of quantum electrodynamics (QED) in the accurate
computation of atomic spectra. However, before addressing this problem, one
must consider first the more fundamental quesion:  is it conceivable that a
model based on the concept of real discrete particles is able to explain
atomic spectra, which are  treated in quantum theory as a pure eigenmode
phenomenon? The answer lies again in the dual nature of the metron model,
which comprises both  field and particle elements. The field content
reproduces the nonlinear wave dynamics of quantum field theory to lowest
interaction order. Thus the spinor and electromagnetic components of the
n-dimensional metric, defined by the polarization relations
(\ref{3.10})\footnote{The specific forms are given in II.}, yield the
coupled field equations of the Maxwell-Dirac system to lowest interaction
order (II). The particle content is an additional feature that complements
but does not alter the wave picture.

\begin{figure}[!t] \centering
$$\epsfig{file=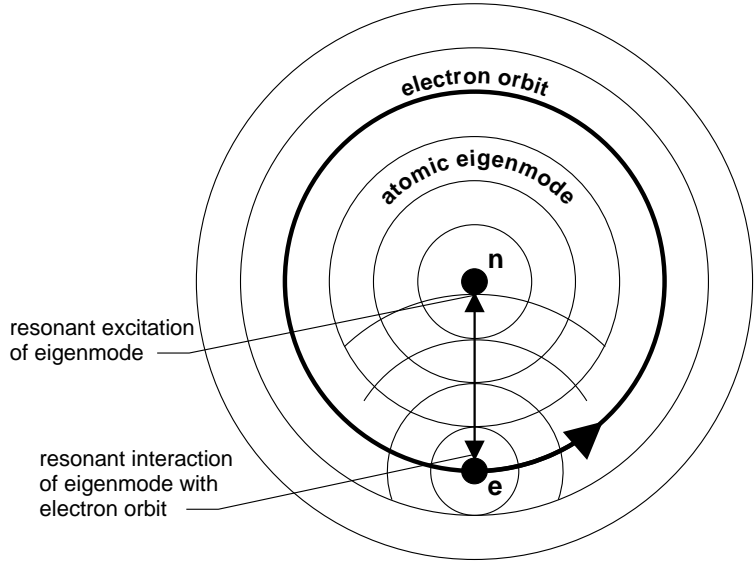, width=10cm}$$
\caption{Resonant interactions between an orbiting electron $e$ and the
Dirac eigenmode generated by the interaction of the electron's de Broglie
field with the atomic nucleus $n$.}
\label{figwolf3}
\end{figure}

For the special case of a Dirac field $\psi$ interacting with the
prescribed electromagnetic field $A_{\lambda}$ of an atomic nucleus, the
field $\psi$ is determined in QED by the Schr\"odinger equation, or in the
general relativistic case, the Dirac equation
\begin{equation} \label{4.20}
D(\psi)  :=   \left  \{
\gamma^{\lambda}   \left (
\partial_{\lambda} - ie A_{\lambda} \right )  +
\omega  \right  \}  \psi = 0.
\end{equation}
This yields a discrete set of trapped eigenmodes $\psi_m$ with
eigenfrequencies $\omega_m$, together with a continuum of free modes. In
the metron model, in which there  exists a real orbiting electron particle,
one is concerned with three fields (Fig.\ref{figwolf3}): the de Broglie
field of the orbiting electron, the electromagnetic field of the atomic
nucleus and the scattered de Broglie field generated by the interaction of
the de Broglie field of the electron with the nucleus. The Dirac field
$\psi$ in (\ref{4.20}) corresponds to the scattered de Broglie field.
However, in  addition to the terms representing the propagation of the
scattered de Broglie field and its interaction with the electromagnetic
field, which are identical to the corresponding terms of the Dirac equation
(\ref{4.20}) of QED, the metron model contains also a forcing term $F$
representing the generation of the scattered field by the orbiting
electron. Thus eq.(\ref{4.20}) is replaced in the metron case by
\begin{equation} \label{4.20a}
D(\psi) =  F
\end{equation}

The forcing function $F$ is periodic with approximately the basic de
Broglie frequency of the orbiting electron, but contains a small frequency
modulation imposed by the electron's orbital motion. If the resulting
forcing frequency is equal to  an eigenfrequency $\omega_m$ of one of the
atomic eigenmodes, the eigenmode is forced in resonance and, in the absence
of further interactions, would grow linearly with time. However, the
eigenmode acts back, again in resonance, on the orbiting electron
(Fig.\ref{figwolf3}). The net effect of the simultaneous resonant
interactions is that the orbiting electron is trapped in discrete orbits
that are in resonance with the eigenmodes of eq.(\ref{4.20}). Under these
resonant trapping conditions, the resonant coupling  generates a mean force
that is able to balance the electromagnetic interactions of the orbiting
electron with a distant ensemble of particles that would otherwise give
rise to radiative damping (III).

For the case of circular orbits, the resonant conditions are found to be
identical to the integer-action conditions of the original Bohr orbital
model. Thus the metron model represents an interesting amalgam of the Bohr
orbital model with QED. Details are given in III.

The question remains whether the metron model is able to reproduce atomic
spectra to the same high level of accuracy as QED. In detail, the metron
interaction computations differ from QED: they contain no closed loop
contributions, there should arise no divergences, so that there is no need
for a renormalization formalism, and the QED and metron interaction
Lagrangians differ at higher order. One can perhaps draw some encouragement
from Barut's (1988) claim that classical tree-level computations of the
eigenmodes of the Dirac equation (\ref{4.20}), without closed loop
contributions and renormalization, yield better agreement with measurements
than the QED computations. However, Barut's computations, although
classical, are also not identical to the metron computations.

I have also not discussed the problem of the transition between eigenmodes
through emitted or absorbed radiation. It is shown in III that the coupling
of atomic eigenmodes through radiation is governed by the same
Maxwell-Dirac field equations in the metron model as in QED. Thus the same
eigenfrequency difference relations and selection rules apply in both
cases. However, the computation of the transition probabilites involve not
only interactions between the eigenmodes and the  emitted or absorbed
radiation, but also changes in the electron orbits, and are more complex in
the metron model than in the QED case.

\section{The Standard Model}
\label{The Standard Model}

The metron derivation of the coupled field equations and particle
distant interactions of the Maxwell-Dirac-Einstein system outlined above
can be readily generalized to the remaining weak and strong interactions.
For the Maxwell-Dirac-Einstein system, the dimension of harmonic space
needed to be at least four to reproduce the 4-spinor Dirac equation (II),
but the periodicities of the metron parton components could be restricted
to time ($k_4$) and -- as in the original Kaluza-Klein model -- a single
harmonic dimension ($k_5$). The wavenumber component $k_5$ determined the
electromagnetic charge, while the frequency $k_4$, or the particle mass,
governed the gravitational coupling. The principal generalization needed to
include the two additional forces is the introduction of further partons
with periodicities in the other harmonic dimensions.

The algebraic details, described in IV, are rather complex and will not be
outlined here. However, the basic geometrical configuration of the metron
model needed to reproduce the principal features of the Standard Model --
the three families of quarks and leptons, together with their coupling
fields, the electroweak bosons and strong-interaction gluons, and the
$U(1)\times SU(2) \times SU(U3)$ gauge symmteries -- can be readily
summarized  (see Fig.~\ref{figwolf4}):

\begin{figure}[!t] \centering
$$\epsfig{file=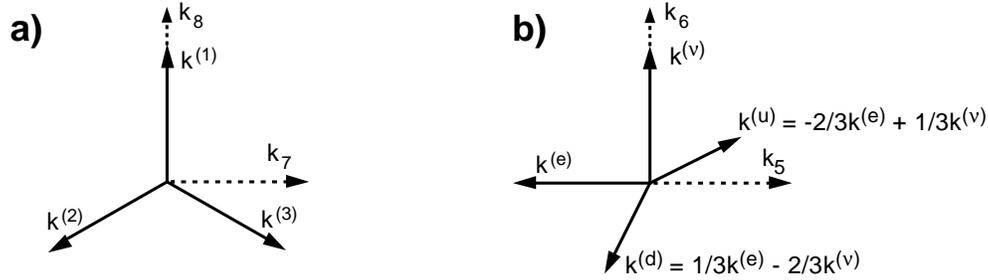, width=13cm}$$
\caption{Fermion harmonic wavenumber configurations for (a) the three
coloured quarks in the chromodynamic plane $k_7,k_8$ and (b) the leptons
$e,\nu$ and quarks $u,d$ in the electroweak plane $k_5,k_6$ (from IV).}
\label{figwolf4}
\end{figure}

\begin{itemize}

\item
To every parton there exists an anti-parton obtained by reflecting the
parton configuration in harmonic space.

\item
Leptons are represented by partons whose  harmonic wavenumber vectors lie
in the electroweak plane $k_5,k_6$. The harmonic wavenumber vectors of the
electron, muon and $\tau$ particle are oriented parallel to the $k_5$ axis,
the wavenumber vectors of their associated neutrinos parallel to the $k_6$
axis.

\item
Quarks are characterized by wavenumber vectors whose major components lie
in the colour plane $k_7,k_8$, but which contain also weak components in
the electroweak plane. The wavenumber components in the colour plane
describe the strong interaction coupling, while the electroweak components
characterize the electroweak splitting between the up/down, charm/strange
and top/bottom quark pairs.

\item
The three Standard Model families are explained as the first three modes
(with respect to the radial coordinate in physical space) of the trapped
metron eigen-oscillations. According to this picture, there should be no
limit to the number of families, although the higher modes presumably
become increasingly unstable.

\item
The three quark colours correspond to three orientations of the quark
wavenumber vectors in the colour plane in the form of a symmetrical
Mercedes star. Quarks occur always in combinations such that the vector sum
of the wavenumber vectors in the colour plane vanishes (the net colour is
white). This condition implies that, in addition to the quadratic
(difference-wavenumber) self-interaction of individual quarks, the joint
interaction of the set of quarks can couple into a mean field (the wave
guide) with zero harmonic wavenumber.

\item
The coupling between quarks is mediated by bosons generated by quadratic
difference interactions between quarks.  There exist also sum- and
higher-order interactions, which have no counterpart in the Standard Model.
Thus the Standard Model appears from the viewpoint of the metron model only
as an approximation.

\item
The gauge symmetries of the Standard Model follow from the geometrical
symmetries of the parton configurations, in combination with the invariance
of the generalized Einstein equations with respect to (n-dimensional)
coordinate transformations. However, there exist minor differences between
the symmetries of the Standard Model and the metron model (for example, the
quark coupling through diagonal and non-diagonal gluons is characterized by
different coupling constants in the metron model). The different origin of
symmetries in the metron model compared with QFT is fundamental. The
Einstein vacuum equations contain no symmetries apart from the invariance
with respect to coordinate transformations and the symmetry of the
background metric (which in the simplest metron model is isotropic in
physical-plus-harmonic space). Thus in contrast to QFT, in which the basic
symmetries are introduced {\it a priori} into the Lagrangians, all
symmetries or asymmetries of the metron model (such as the different role
of physical spacetime and harmonic space) are properties of the solutions
rather than the field equations themselves.

\item
All 23 empirical constants of the Standard Model follow from the
geometrical structure of the metron solutions.
\end{itemize}

\section{Conclusions}
\label{Conclusions}

As pointed out already by Kaluza and Klein for the case of
electromagnetism, the extension of the classical gravitational equations to
a higher dimensional space is the simplest way to generalize Einstein's
elegant concept of gravity to other forces. The reduction to the vacuum
equations (\ref{w6}) simplifies the equations still further. More
importantly, this step opens the possibility of explaining rather than
postulating the existence of particles and their interactions.

However, a prerequisite for pursuing this avenue is the resolution of the
wave-particle duality paradox. This can be achieved within the framework of
classical physics, based on the objective existence of both fields and
particles, by assuming that particles carry periodic far fields
characterized by de Broglie's dispersion relation. Bell's theorem that such
hidden-variable theories are necessarily in conflict with the
Einstein-Podolsky-Rosen experiment is circumvented by the observation that
at the micro-physical level strict time-reversal symmetry applies, as
opposed to the existence of an arrow-of-time assumed by Bell. An
arrow-of-time arises only at the macrophysical level of irreversible
phenomena.

It is rather gratifying that a theory based on these simple concepts is
able to reproduce the coupled field equations of quantum field theory at
lowest interaction order. In addition, the theory explains the origin and
derives the structure of discrete particles, including the internal forces
within the particle core and the distant interaction forces of gravity and
electromagnetism. In the process, one recovers all fundamental physical
constants: Planck's constant, the gravitational coupling, the particle
masses and charges, and all other empirical constants of the Standard
Model. The origin of the extreme weakness of the gravitational forces is
also explained.

However, the metron model outlined in this brief sketch, and developed more
fully in the detailed analysis of I-IV, represents still only a skeleton.
Plausibility arguments are given in I for the existence of metron solutions
with the postulated properties needed to recover (and explain) the
principal features of quantum field theory and the elementary particle
spectrum as outlined in II-IV. Explicit computations of coupled
wave-mode/wave-guide metron-type solutions were furthermore presented in I
for a simplified scalar Lagrangian that mirrors the principal nonlinear
properties of the Einstein vacuum equations. However, metron solutions for
the real n-dimensional gravitational Lagrangian still need to be computed.
Work is currently in progress on the numerical determination of metron
solutions for n=8.

It is hoped that numerical computations combined with further theoretical
analysis will help answer a number of questions that have appeared at the
present level of investigation:
\begin{enumerate}
\item
Are metron solutions, assuming they exist, stable?
\item
The metron-type solutions computed in I for a scalar Lagrangian represent a
continuum rather than a discrete spectrum. How can one explain the
existence of a discrete rather than a continuous spectrum of elementary
particles? And why do the solutions exhibit the  particular structure that
was postulated, in particular a periodicity in one sub-space and locality
in another? Is this related to the first question, namely a stability
criterion?
\item
Although the scattering of particles at a lattice or passing through a slit
screen yields qualitatively similar interference patterns in both the
metron model and quantum theory, the scattering computations for the two
theories probably differ in detail. Can one distinguish experimentally
between the two theories?\footnote{This is not as straightforward as it may
seem, as de Broglie scattering computations are normally carried out in the
inverse mode: the interaction potential is inferred from the interference
pattern.}
\item
Are higher-order metron computations of atomic spectra consistent with
experiment and the highly accurate results of quantum electrodynamics?
\item
Does the metron model yield the correct transition probabilities for the
atomic emission and absorption of radiation?
\item
Does the metron model reproduce the Bohr orbital picture not only for
circular but also elliptical orbits?
\end{enumerate}
Clearly, there are still many fundamental questions to be answered.
However, the ability to reproduce and explain a broad spectrum of basic
microphysical phenomena within the framework of a unified classical
``objective'' physical theory, starting only from the deceptively innocuous
Einstein vacuum equations $R_{LM}=0$ in an 8-dimensional space, is at least
an encouraging first step towards realizing Einstein's dream of the
unification of all forces of nature in a single deterministic theory.

\subsection*{Acknowledgements}
It is pleasure to express my thanks to Wolfgang Kundt for many fruitful
discussions and constructive comments on the first draft of this paper, and
for the opportunity to present these ideas on the occasion of his highly
enjoyable 65'th birthday symposium.

\end{document}